\documentclass[aps,prd,nofootinbib,showpacs,superscriptaddress,floatfix]{revtex4}
\usepackage{amssymb,amsfonts}
\usepackage{bm} 
\usepackage{graphicx}

\newcommand{\der}[2]{\frac{\partial #1}{\partial #2}}

\newcommand{\dder}[2]{\frac{\partial^2 #1}{\partial #2 ^2}}
\newcommand{\w}[1]{\bm{#1}}
\newcommand{\be}{\begin{equation}}
\newcommand{\ee}{\end{equation}}
\newcommand{\bea}{\begin{eqnarray}}
\newcommand{\eea}{\end{eqnarray}}

\newcommand{\M}{{\mathcal M}}
\newcommand{\Sp}{{\mathcal S}}

\newcommand{\Liec}[1]{{\mathcal L}_{\w{#1}}\,}

\newcommand{\DSc}{\mathcal{D}}

\newcommand{\tD}{\tilde D}
\newcommand{\lapang}{\Delta_{\theta\varphi}}

\begin{document}

\title
{Mathematical Issues in a Fully-Constrained Formulation of Einstein Equations}

\newcommand*{\VAL}{Departamento de Astronom\'\i a y Astrof\'\i sica, Universidad de 
Valencia, Valencia, Spain}
\newcommand*{\IAA}{Instituto de Astrof\'{\i}sica de Andaluc\'{\i}a, CSIC, Apartado Postal 
3004, Granada 18080, Spain} 
\newcommand*{\MEU}{Laboratoire Univers et Th\'eories (LUTH), Observatoire de Paris, CNRS, 
Universit\'e Paris Diderot, Place Jules Janssen, 92190 Meudon, France} 

\author{Isabel Cordero-Carri\'on}
\email{Isabel.Cordero@uv.es}
\affiliation{\VAL}

\author{Jos\'e Mar\'\i a Ib\'a\~nez}
\email{Jose.M.Ibanez@uv.es}
\affiliation{\VAL}

\author{Eric Gourgoulhon}
\email{eric.gourgoulhon@obspm.fr}
\affiliation{\MEU}

\author{Jos\'e Luis Jaramillo}
\email{jarama@iaa.es}
\affiliation{\IAA}\affiliation{\MEU}

\author{J\'er\^ome Novak}
\email{Jerome.Novak@obspm.fr}
\affiliation{\MEU}

\date{18 February 2008}

\begin{abstract}

  Bonazzola, Gourgoulhon, Grandcl\'ement, and Novak [Phys. Rev. D {\bf 70}, 104007 (2004)]
  proposed   a new formulation for 3+1 numerical relativity. Einstein equations result,
  according to that formalism, 
  in a coupled elliptic-hyperbolic system. We have carried out a preliminary
  analysis of the mathematical structure of that system, in particular
  focusing on the equations governing the evolution for the deviation of a
  conformal metric from a flat fiducial one. The choice of a Dirac's gauge for
  the spatial coordinates guarantees the mathematical characterization of that
  system as a (strongly) hyperbolic system of conservation laws. In the
  presence of boundaries, this characterization also depends on the boundary
  conditions for the shift vector in the elliptic subsystem. This interplay
  between the hyperbolic and elliptic parts of the complete evolution system
  is used to assess the prescription of inner boundary conditions for the
  hyperbolic part when using an excision approach to black hole spacetime
  evolutions.
\end{abstract}


\maketitle


\section{A Fully-Constrained evolution scheme}
\label{s:Introduction}
A second-order fully-constrained evolution formalism for the Einstein
equations has been proposed in Ref.~\cite{BonazGGN04}.  This evolution scheme,
that will be referred in the following as Fully-Constrained Formulation (FCF),
is based on a conformal 3+1 formulation of General Relativity and makes use of
an elliptic condition for the choice of spatial coordinates, a {\em
generalized Dirac gauge}, and a maximal condition for the slicing.  The
enforcement of the constraints along the evolution together with the elliptic
nature of the employed gauge conditions, translates the FCF formalism into a
mixed elliptic-hyperbolic Partial Differential Equations (PDE) system,
consisting in five quasi-linear elliptic equations coupled with a tensorial
second-order in time and in space evolution equation for the conformal metric.
In this article, we aim at gaining insight on some mathematical issues
associated with this PDE system and, in particular, assessing the
hyperbolicity of the tensorial evolution part.  A good understanding of the
mathematical structure of the system will be crucial in the context of full 3D
numerical relativity simulations, since the choice of state-of-the-art
numerical tools will be adapted to the specific structures of the whole system
governing the evolution of matter fields in a dynamical space-time: spectral
methods for the elliptic subsystem \cite{GrandN07}, and modern high-resolution
shock-capturing techniques for the hyperbolic part \cite{RevMarti,RevFont}.
The implementation of the scheme in \cite{BonazGGN04} will naturally extend
previous works ---following the Conformal Flatness Condition (CFC) approach of
Isenberg-Wilson-Mathews \cite{Isenberg,Wilson96}--- devoted to the study of
some relevant astrophysical sources of gravitational radiation
\cite{DFM01,DFM02a,DFM02b,MDM05}.

\subsection{Gauge reduction, PDE evolution systems and
well-posedness}
The gauge character of General Relativity (GR)  
strongly conditions any attempt of finding a solution
by solving a Partial Differential Equations 
(PDE) problem.
In its standard formulation through the Einstein equation
\be
\label{e:EFE}
R_{\mu\nu}-\frac{1}{2}R \;g_{\mu\nu}= 8\pi \;T_{\mu\nu}, 
\ee 
solutions are given in terms of spacetime geometries $({\cal M}, g_{\mu\nu})$, 
i.e. classes of Lorentzian metrics $  g_{\mu\nu}$
equivalent under diffeomorphisms of ${\cal M}$, rather than 
by specific 4-metrics in some particular coordinate system.
As a consequence of this, any attempt to cast (\ref{e:EFE}) 
as a standard PDE system necessarily must go through a 
{\em gauge reduction} process.
This fixing of the gauge involves four different (differential)
systems: i) the {\em reduced} system, whose solution provides
the metric in a given coordinate system, ii) the {\em constraint system},
consequence of the gauge character of the theory and that characterizes
the solution manifold, 
iii) the {\em gauge system}, which fixes the coordinate chart and
permits to write the reduced system as a standard PDE problem, and
iv) the {\em subsidiary system}, guaranteeing the overall consistency
along the evolution and, in particular, between 
the reduced and gauge systems.
The mathematical consistency of the evolution formalism involves
two aspects. First, one must assess the analytic well-posedness 
of the PDE system that is actually solved during the evolution,
that we will refer to in the following as the {\em evolution PDE system}, that includes 
the {\em reduced system} but possibly other additional PDEs.
Second, one must guarantee the fulfillment of the {\em subsidiary system} during
the evolution.

As in other evolution formalisms based on the Initial Value
problem for the Einstein equation \cite{FriRen00}, 
the constrained system in the FCF scheme follows from the
Gauss-Codazzi-Ricci conditions
\bea
\label{e:constraints}
{}^{(3)}\!R - K_{ij}K^{ij} + K^2 &=& 16 \pi \rho \nonumber \\
D_j\left(K^{ij} -\gamma^{ij}K\right) &=& 8\pi J^i  \ \ ,
\eea
i.e. the Hamiltonian and momentum constraints in the 3+1 formulation
($\rho$ is the energy density and
$J^i$ the current vector) which are elliptic in nature.
The currently most successful numerical evolution formalisms are {\em free} 
schemes in which the constraint 
system (\ref{e:constraints}) is not enforced during the evolution.
This is the case of
certain generalized harmonic formalisms \cite{Pre05b,GunMarCal05}
and the 3+1 BSSN (from Baumgarte, Shapiro, Shibata and Nakamura; see references \cite{ShiNak95,BauSha98}) used in recent binary black hole 
breakthroughs \cite{Pre05,CamLouMar06,BakCenCho06,Pre07} and in  
fully 3D evolution of binary neutron 
stars (see e.g. \cite{ShiUry00}).
In these free schemes, the corresponding evolution PDE system is formed by the
respective reduced systems together with some additional evolution
equations to fix the harmonic gauge sources, in the case generalized
harmonic schemes, or the lapse function and shift vector, in the BSSN case.
No elliptic equation is solved during the
evolution and standard hyperbolic techniques can in principle be
used to assess the well-posedness of the evolution system (cf. in this
sense \cite{GunMar06b} for the case of the BSSN system). 
In contrast, the FCF here discussed actually incorporates the constraints
to the evolution PDE system. Moreover, the use of the above-mentioned 
elliptic gauge conditions adds additional elliptic equations
during the evolution. The resulting FCF scheme presents some interesting properties as compared
with free evolution schemes. Apart from the absence of constraint violations
(an issue under control in current BSSN and generalized harmonic formulations),
we can highlight the following features (cf. \cite{BonazGGN04} for a more complete discussion):
first, the FCF naturally generalizes (as commented above) the successful scheme
employed in the CFC approximation to 
General Relativity; second, it permits to read the gravitational waveforms directly from the 
metric components; third, the scheme can be straightforwardly
adapted to the extraction of gravitational radiation at null infinity by 
making use of hyperboloidal 3-slices implemented by means of a 
{\em constant mean curvature} elliptic gauge condition; and fourth, it provides
a well-suited framework for the formulation of realistic (approximate)
prescriptions in the construction of
quasi-stationary astrophysically configurations \cite{UryLimFri06}.
However, the well-posedness analysis of such a mixed elliptic-hyperbolic system can
be a formidable problem, since part of the dynamics related
to the characteristic fields in the hyperbolic part is encoded 
in fields obtained only once the elliptic part is solved. 
Even though analyses of such systems
exist in the GR literature (see e.g. 
Refs.\cite{ChoYor95,ChoYor96,ChoYor96b} and
particularly Ref. \cite{AndMon03}) they deal with free 
evolution systems, in which the elliptic part follows
only from the gauge conditions.
The well-posedness analysis of the complete elliptic-hyperbolic system 
in the FCF scheme, which in addition includes the constraints, is beyond the scope
of this work and we will mainly focus on the hyperbolicity
analysis of the tensorial evolution equation.  
Before referring to the additional issues related to the subsidiary system,
we must provide some details about the FCF formalism.

\subsection{Brief review of the FCF scheme}\label{ss:brief_review}
Following Ref.~\cite{BonazGGN04}, we consider a standard 3+1 decomposition of
an asymptotically flat spacetime $(\M, g_{\mu\nu})$ in terms of a foliation by
spacelike hypersurfaces $(\Sigma_t)$.  We denote the unit timelike normal
vector to the spacelike slice $\Sigma_t$ by $n^\mu$, the spatial 3-metric by
$\gamma_{\mu\nu}$, i.e. $\gamma_{\mu\nu}= g_{\mu\nu}+n_\mu n_\nu$, and
adopt the following sign convention for the extrinsic curvature:
$K_{\mu\nu}=-\frac{1}{2}{\cal L}_{\w n}\gamma_{\mu\nu}$.  The evolution vector
$t^\mu\equiv (\partial_t)^\mu$ is decomposed in terms of the lapse function
$N$ and the shift vector $\beta^\mu$, as $t^\mu= N n^\mu+\beta^\mu$.

Under this 3+1 decomposition, Einstein equation (\ref{e:EFE}) splits
into the 3+1 constraints in (\ref{e:constraints}) and a set
of evolution equations for the extrinsic curvature that,
together with the kinematical relation defining the extrinsic curvature,
constitute the 3+1 evolution equations 
\bea
&&\left(\partial_t - {\cal L}_\beta\right)\gamma_{ij}= -2N K_{ij}   \nonumber
 \\
&&\left(\partial_t - {\cal L}_\beta\right) K_{ij} = 
-D_iD_jN + 
N\left\{{}^{(3)}\!R_{ij}+ K K_{ij} - 2{K_i}^k K_{kj}
+4\pi\left[(S-E)\gamma_{ij} - 2S_{ij}\right]\right\} \label{e:ADMsystem}  \ \ .
\eea
This is a first-order in time and second-order in space evolution system
for $(\gamma_{ij},K^{ij})$.

The first specific element in the FCF scheme is the introduction of a {\em
  time independent} fiducial flat metric $f_{ij}$, which satisfies ${\cal
  L}_{\w{t}}f_{ij}=\partial_t f_{ij}=0$. This rigid structure is chosen to
coincide with $\gamma_{ij}$ at spatial infinity, capturing its asymptotic
Euclidean character, and permits to work with tensor quantities rather than
with tensor densities.  We will denote by ${\cal D}_i$ the Levi-Civita
connection associated with $f_{ij}$.

\paragraph{Conformal decomposition.}
As a step forward in the reduction process to the PDE system in the present FCF, 
we perform a conformal decomposition of the 3+1 fields: 
\bea
\label{e:conformal_decomp}
\gamma_ {ij} =\Psi^4\tilde{\gamma}_{ij}  \ \ , 
K^{ij} =\Psi^4\tilde{A}^{ij}+\frac{1}{3}K\gamma^{ij} \ \ ,
\eea
where $K=\gamma^{ij}K_{ij}$, the representative $\tilde{\gamma}_{ij}$ 
of the conformal class of the 3-metric is chosen to satisfy the 
unimodular condition
$\mathrm{det}(\tilde{\gamma}_{ij})=\mathrm{det}(f_{ij})$,
and the traceless part $\tilde{A}^{ij}$ of the extrinsic curvature is decomposed as
\be
\label{e:Aij}
\displaystyle{\tilde{A}^{ij} = \frac{1}{2N}
        \left( \tilde{D}^i\beta^j + \tilde{D}^j\beta^i
	         - \frac{2}{3}  \tilde{D}_k\beta^k \tilde{\gamma}^{ij}
		          + \partial_t{\tilde\gamma}^{ij}  \right)} ,
\ee
with $\tilde{D}_i$ the Levi-Civita connection associated 
with $\tilde{\gamma}_{ij}$. Finally, in the following we will denote
by $h^{ij}$ the deviation of the conformal metric from the flat fiducial
metric, i.e.  \be
\label{e:hij}
h^{ij} := {\tilde{\gamma}}^{ij} - f^{ij}.
\ee
Using these conformal decompositions of $\gamma_{ij}$ and $K^{ij}$,
the 3+1 constraints (\ref{e:constraints}) and evolution system 
(\ref{e:ADMsystem}) can be expressed in terms of the basic variables
$h^{ij}, \Psi, N, \beta^i, K$. Before giving more explicit expressions,
let us remove the gauge freedom.

\paragraph{Gauge system.} 
Following the prescriptions in \cite{BonazGGN04}, namely maximal
slicing and the so-called generalized Dirac gauge, we choose
\be
\label{e:gauges}
K=0, \ \ H^i := \DSc_k\tilde{\gamma}^{ki}=0,
\ee
These gauge conditions fix the coordinates, even in the initial slice,
up to boundary terms (see e.g. sections 9.3. and 9.4. in \cite{Gou07a}).
These two relations define the gauge system in the FCF scheme. 
Since the gauge system is meant to hold at all times, the following conditions must also be satisfied
\be
\label{e:gauges_evol}
\dot{K}=0, \ \ \partial_t\left({\DSc_k\tilde{\gamma}^{ki}}\right)=0 .
\ee
The FCF scheme actually enforces the first of these
conditions, $\dot{K}=0$, during the evolution. Taking the trace
in the second equation in (\ref{e:ADMsystem}), and using the Hamiltonian constraint that is also enforced during the evolution
(see below), an elliptic equation for the lapse follows
\be
\label{e:N_equation}
\displaystyle \tD_k \tD^k N + 2 \tD_k\ln\Psi\, \tD^k N
	            = 
		    S_{N}[N, \Psi, \beta^i,\tilde{\gamma}_{ij}].
\ee
\paragraph{Main or reduced system.}
In the FCF scheme in Ref. \cite{BonazGGN04} the reduced system
is a second-order in time and second-order in space 
evolution system for the deviation tensor $h^{ij}$. This is obtained
by: i) combining equations in (\ref{e:ADMsystem})  
into a single second-order in time equation; ii) inserting in it 
the conformal decompositions  
(\ref{e:conformal_decomp}) and (\ref{e:Aij}), and iii) imposing the
gauges (\ref{e:gauges}). The resulting expression is formally written
as (see next section for a detailed account):
\be
\label{e:hyperbolic_part}
\frac{\partial^2 h^{ij}}{\partial t^2} - \frac{N^2}{\Psi^4}
\tilde{\gamma}^{kl}{\cal D}_k {\cal D}_l
h^{ij}
- 2 {\cal L}_{\beta}\frac{\partial h^{ij}}{\partial t} + {\cal
  L}_{\beta}{\cal L}_{\beta}h^{ij} = S_{h}^{ij},
\ee
where the source $S_{h}^{ij}$ does not contain second derivatives
of $h^{ij}$.
Use of the Dirac gauge results in the wave-like form of this equation, since it
eliminates certain second derivatives of the type ${\cal D}^i {\cal D}_k h^{kj}$
coming from the expression of the Ricci tensor.
\paragraph{Constrained system.} 
The Hamiltonian constraint in (\ref{e:constraints}) can be written as an elliptic equation 
for the conformal factor $\Psi$:
\be
\label{e:conformal_factor}
 \tD_k \tD^k \Psi - \frac{{}^3\!{\tilde R}}{8} \, \Psi =
  S_{\Psi}[\Psi, N, \beta^i, \tilde{\gamma}_{ij}].
\ee
Again $S_{\Psi}[\Psi, N, \beta^i, \tilde{\gamma}_{ij}]$ represents
a non-linear source.
Momentum constraint poses a more subtle issue.
In Ref. \cite{BonazGGN04} an elliptic equation for the shift vector
is deduced using {\em both} the momentum constraint and the 
preservation in time of the Dirac gauge (second relation in 
(\ref{e:gauges_evol})):
\be
\label{e:shift}
      \displaystyle \tD_k \tD^k \beta^i +  \frac{1}{3} \tD^i \tD_k \beta^k
          + {}^3\!{\tilde R}^i_{\ \, k} \beta^k  =  S^i_{\beta}[\Psi,
	   N, \beta^i, \tilde{\gamma}_{ij}]   
\ee 
An equation for the shift could be derived from the momentum constraint alone,
but the coupling to the tensorial equation (\ref{e:hyperbolic_part}) would become 
more complicated due to the presence of a mixed time-space 
second-order derivative of $h^{ij}$. This term is eliminated by the use 
of a Dirac,  or a similar, gauge. 

Alternatively, an elliptic equation for the shift can be 
drawn from the preservation of the Dirac gauge alone, renouncing, therefore, to the fully-constrained character of the scheme
---e.g. this is the strategy in Ref. \cite{AndMon03}, but using a spatial harmonic
gauge condition instead of the Dirac one.
At the end of the day,
the choice (\ref{e:shift}) in the FCF scheme provides
an elliptic equation for the shift that enforces the momentum constraint,
{\em as long as the Dirac gauge is satisfied}.

\paragraph{FCF evolution PDE system.}
The mixed elliptic-hyperbolic PDE system
that evolves some initial data given on an Cauchy slice is
formed by: a) Eqs. (\ref{e:N_equation}), (\ref{e:conformal_factor}) and
(\ref{e:shift}), the elliptic part, and b) Eq. (\ref{e:hyperbolic_part}), the wave-like tensorial equation.  As we have
pointed out, we will not consider here the well-posedness analysis of the
whole system. To give an idea of the involved difficulties, we note that the
elliptic part is very similar to the Extended Conformal Thin Sandwich (XCTS)
\cite{PfeYor03,Pfe05} employed in the construction of initial data, though here it is
solved all along the evolution. Even the restriction to the elliptic subsystem
represents a very hard problem, as it is illustrated by the lack of the
existence results for the XCTS system and the preliminary numerical \cite{PfeYor05} (see
also \cite{JarAnsLim07})
and analytical \cite{BauOMuPfe07,Wal07} results pointing towards a generic non-uniqueness
of the elliptic system.  For these reasons, we will focus on the study of the
hyperbolicity of the tensorial evolution equation (\ref{e:hyperbolic_part}), 
understanding this as a
necessary condition for the overall well-posedness.

\paragraph{Subsidiary system.}
The resolution of the PDE evolution system only guarantees the consistency
between the reduced and gauge systems as far as the slicing condition is
regarded, since equation (\ref{e:N_equation}) for the lapse is indeed
enforced. This is in principle not the case for the Dirac gauge. More
dramatically, if the Dirac gauge is actually not satisfied, the FCF scheme is
not really fully-constrained, since in that situation Eq. (\ref{e:shift}) no
longer enforces the momentum constraint. A control of the evolution of the
Dirac gauge is therefore crucial in the scheme.  A wave-like equation for
${\cal D}_k h^{ki}$ can be obtained by taking the divergence of the tensorial
Eq. (\ref{e:hyperbolic_part}). The vanishing of ${\cal D}_k h^{ki}$ in the
evolution would then follow from the initial conditions $ {\cal D}_k h^{ki}=0$
and $\partial_t\left( {\cal D}_k h^{ki}=0\right)=0$ imposed in the
construction of the initial data, and the satisfaction of Eq. (91) in Ref.
\cite{BonazGGN04} for $\dot{\beta^i}$.  The latter can be considered as the
{\em subsidiary system} in the FCF scheme.

\subsection{Specific objectives and organization}
Though the wave character of Eq. (\ref{e:hyperbolic_part}) essentially guarantees
its hyperbolicity, we aim here at developing a more detailed analysis.
This is motivated by the need of controlling the characteristics in initial boundary problems 
and also when trying to make use of first-order techniques employed in matter evolutions.
Our main specific goal in this article is the development of a hyperbolicity
analysis of a {\em first-order} version of the evolution part in the FCF formalism, where $N$,
$\Psi$ and $\beta^i$ are considered as fixed parameters. In particular, we aim
at obtaining explicit expressions for the characteristic fields and speeds.
As pointed out above, this point represents a fundamental ingredient in the study of the appropriate
boundary conditions if boundaries are present in the integration domain. This
constitutes only a preliminary study of the well-posedness of the evolution
system since no stability analysis whatsoever will be considered. Certainly
further analysis is required.  However, in the absence of a full treatment and
being ultimately motivated by practical numerical implementations
needs, the level of rigor and completeness in this article is adapted to the
achievement of limited but concrete results.

On behalf of self-consistency, and in spite of the lack of a fully rigorous
treatment of the FCF subsidiary system, we also aim at discussing certain
(numerical) algorithms devised to guarantee the fulfillment of the Dirac gauge
along the evolution. Though this is not the substitute of a formal proof
it provides, on the one hand,  
support for the coherence among the reduced, gauge and constrained
systems. On the other hand, and more importantly from a practical point of view,
the implementation of the FCF scheme is then guaranteed to be fully-constrained, even
in numerical implementations where errors can occur even if analytic
well-posedness has been established.

The article is organized as follows. Section \ref{s:first_order_reduction}
presents first-order formulation of the FCF scheme, more concretely of its
reduced system.  In section \ref{s:characteristics} the characteristic
structure of the reduced system is analyzed, with a brief application to inner
boundaries in excised black hole spacetime evolutions. Section
\ref{s:conservation_laws} discusses the possibility of writing the first-order
reduced FCF system as a system of conservation laws, by making explicit use of
the Dirac gauge.  In section \ref{s:Dirac_system} two different manners of
enforcing the Dirac gauge in the evolution are introduced, providing key
support for overall consistency and guaranteeing the fully-constrained
character of the scheme.  Finally section \ref{s:discussion} concludes with a
discussion of the results.

\section{First-order reduction of the reduced system 
in the FCF}
\label{s:first_order_reduction}
Equations governing the evolution of $h^{ij}$ in the FCF are:
\bea
\label{e:hyperbolic_part_detailed}
&&\displaystyle{\frac{\partial^2h^{ij}}{\partial t^2} - \frac{N^2}{\psi^4}\tilde{\gamma}^{kl} \DSc_k \DSc_l h^{ij}
	- 2{\Liec{\beta}} \frac{\partial h^{ij}}{\partial t} + {\Liec{\beta}}{\Liec{\beta}}h^{ij}}
	\nonumber \\ 
&&	= {\Liec{\dot{\beta}}}h^{ij}
	+\displaystyle{ \frac{4}{3}{\DSc}_k\beta^k\left(\frac{\partial}{\partial t}
	- {\Liec{\beta}}\right)h^{ij}	} 
	\nonumber \\ 
&&	-\displaystyle{\frac{N}{\psi^6} \DSc_k Q\left(\DSc^ih^{jk}
	+ \DSc^j h^{ik} - \DSc^k h^{ij}\right) }
	\nonumber\\
&& + \displaystyle{\left[\left(\frac{\partial}{\partial t} - {\Liec{\beta}}\right)ln N\right]
	\left[\left(\frac{\partial}{\partial t} - {\Liec{\beta}}\right)h^{ij} \right.}
	\nonumber\\
&&	- \left. \frac{2}{3}\DSc_k\beta^kh^{ij} 
	+ \left(L\beta\right)^{ij}\right]
	\nonumber\\
&&      +\displaystyle{ \frac{2}{3}\left[\left(\frac{\partial}{\partial t} - {\Liec{\beta}}\right)\DSc_k\beta^k
	- \frac{2}{3}\left(\DSc_k\beta^k\right)^2\right]h^{ij}}
	\nonumber\\
&&	- \left(\displaystyle{\frac{\partial}{\partial t}} - {\Liec{\beta}}\right)\left(L\beta\right)^{ij}
	+\displaystyle{ \frac{2}{3}\DSc_k\beta^k\left(L\beta\right)^{ij}} 
	\nonumber\\
&&	+ 2N\psi^{-4}Z^{ij} 
	\nonumber\\
&&      + \left(2N\right)^2\left[\tilde{\gamma}_{kl}A^{ik}A^{jl} - 4\pi\left(\psi^4S^{ij}
	- \displaystyle{\frac{1}{3}}S\tilde{\gamma}^{ij}\right)\right]
	\nonumber\\
&&	- 2N\psi^{-6} 
	 \left[\tilde{\gamma}^{ik}\tilde{\gamma}^{jl}\DSc_k\DSc_lQ
	+\displaystyle{ \frac{1}{2}}\left(h^{ik} \DSc_l h^{lj} \right. \right.
	\nonumber\\
&&	\left. \left.  +  h^{jk} \DSc_k h^{il} - h^{kl} \DSc_k h^{ij}\right)
	\DSc_lQ - \displaystyle{\frac{1}{3}}\tilde{\gamma}^{ij}\tilde{\gamma}^{kl}\DSc_k\DSc_lQ\right], \nonumber\\
\eea
where $S^{ij}$ and $S$ are, respectively, the spatial components of the stress tensor 
$S_{\alpha\beta}:=\gamma_{\alpha}^{\mu}\gamma_{\beta}^{\nu}T_{\mu\nu}$, associated with the matter energy-momentum tensor $T_{\mu\nu}$, and its trace.
$(L\beta)^{ij}$ 
is the conformal Killing operator associated with the flat metric $f_{ij}$ acting
on the vector field $\beta^i$:
\be
  \left(L\beta\right)^{ij}:=\DSc^i\beta^j + \DSc^j\beta^i - \frac{2}{3}\DSc_k\beta^kf^{ij} \ \ ,
\ee
and the auxiliary quantities $Q$ and $Z^{ij}$ are
\be
	Q:=N\psi^2 \,\,\,,
\ee
\bea
	Z^{ij} & = & N\left[\tilde{R}_*^{ij} + 8\psi^{-2}\left(\tilde{\gamma}^{ik}\DSc_k\psi\right)
	\left(\tilde{\gamma}^{jl}\DSc_l\psi\right)\right]
	 \nonumber\\
	& & + 4\psi^{-1}\left(\tilde{\gamma}^{ik}\DSc_k\psi\right)\left(\tilde{\gamma}^{jl}\DSc_lN\right)
	 \nonumber\\
	& & + 4\psi^{-1}\left(\tilde{\gamma}^{jk}\DSc_k\psi\right)\left(\tilde{\gamma}^{il}\DSc_lN\right)
	 \nonumber\\
	& & - \frac{1}{3}N\left[\tilde{R}_*
	+ 8\psi^{-2}\DSc_k\psi\left(\tilde{\gamma}^{kl}\DSc_l\psi\right)\right]\tilde{\gamma}^{ij}
	 \nonumber\\
	& & - \frac{8}{3}\psi^{-1}\DSc_k\psi\left(\tilde{\gamma}^{kl}\DSc_kN\right)\tilde{\gamma}^{ij}.
\eea
The symmetric tensor $\tilde{R}_*^{ij}$ is defined by
\bea
\label{e:hyperbolic_part_detailed2}
	\tilde{R}_*^{ij} & := & \frac{1}{2}\left[-\DSc_l h^{ik}\DSc_k h^{jl}
	- \tilde{\gamma}_{kl}\tilde{\gamma}^{mn}\DSc_m h^{ik}\DSc_n h^{jl} \right.
	 \nonumber\\
	& & \left. + \tilde{\gamma}^{nl}\DSc_k h^{mn}
	\left(\tilde{\gamma}^{ik}\DSc_m h^{jl}
	+ \tilde{\gamma}^{jk}\DSc_m h^{il}\right)\right] 
	\nonumber\\
	& & + \frac{1}{4}\tilde{\gamma}^{ik}\tilde{\gamma}^{jl}\DSc_k h^{mn}
\DSc_l\tilde{\gamma}_{mn} \ \ ,
\eea
and the scalar $\tilde{R}_*$ is
\bea
  \tilde{R}_*:=\frac{1}{4}\tilde{\gamma}^{kl}\DSc_k h^{mn}\DSc_l\tilde{\gamma}_{mn} - \frac{1}{2}\tilde{\gamma}^{kl}\DSc_k h^{mn}\DSc_n\tilde{\gamma}_{mn} \ \ .
\eea
Let us write Eqs. (\ref{e:hyperbolic_part_detailed})
as a first-order system, by introducing the following auxiliary variables:
\be
\label{u}
	u^{ij}:=\frac{\partial h^{ij}}{\partial t},
\ee
\be
\label{w}
	w^{ij}_{k}:=\DSc_kh^{ij}.
\ee
With these new variables the system for $h^{ij}$ can be cast into
\bea
\label{dudt}
\lefteqn{ \frac{\partial u^{ij}}{\partial t} - \frac{N^2}{\psi^4}\tilde{\gamma}^{kl}\DSc_kw^{ij}_l
  - 2\beta^k\DSc_ku^{ij} + \beta^k\beta^l\DSc_kw^{ij}_l  } \nonumber\\
 & & = \phi^{ij}
	\left(\beta^k,N,\psi,\partial_{\mu}\beta^k,\partial_{\mu}N,\partial_{\mu}\psi,h^{ij},u^{ij},w^{ij}_k\right),
\eea
where $\phi^{ij}$ are source terms which do not contain partial derivatives of 
$u^{ij}$ or $w^{ij}_k$.
From definition $\left(\ref{w}\right)$ we obtain
\be
\label{dwdt}
  \frac{\partial w^{ij}_k}{\partial t} = \DSc_ku^{ij} \ \ ,
\ee
\noindent
where we have taken into account that $\partial_tf^{ij}=0$.               
In terms of the above new auxiliary variables, the system of Eqs. 
$\left(\ref{u},\ref{dudt},\ref{dwdt}\right)$, can be written as:
\be
  \frac{\partial \mathrm{{\bf \bar{v}}}}{\partial t} + \mathrm{ {\bf A}}^l \DSc_l \mathrm{ {\bf \bar{v}} = 
  {\bf g}}\left( \beta^k,N,\psi,\partial_{\mu}\beta^k,\partial_{\mu}N,\partial_{\mu}\psi,h^{ij},u^{ij},w^{ij}_k\right),
  \label{sisth1}
\ee
where the vector $\mathrm{{\bf\bar{v}}}$ is:
\be
\label{bar_v}
  \mathrm{{\bf\bar{v}}} = \left( \begin{array}{l} \left(h^{ij}\right)\\ \left(u^{ij}\right)\\
  \left(w^{ij}_k\right) \end{array} \right),\;\;\; 
\ee
\noindent
and the source $\mathrm{{\bf g}}$ is
\be
  \mathrm{{\bf g}}\left(
  \beta^k,N,\psi,\partial_{\mu}\beta^k,\partial_{\mu}N,\partial_{\mu}\psi,h^{ij},u^{ij},w^{ij}_k\right)
  = \left( \begin{array}{l} \left(u^{ij}\right)\\ \left(\phi^{ij}\right)\\ \left(0\right) \end{array} \right).
\ee
\noindent
In these equations, $\mathrm{{\bf\bar{v}}}$ and $\mathrm{{\bf g}}$ are vectors of dimension 30, as it
results from the symmetry properties of $h^{ij}$, $u^{ij}$, and $w^{ij}_k$.
Let us remind that, besides the above symmetry properties, the following
algebraic constraints have to be satisfied: i) $\rm det\,\, \it
\tilde{\gamma}_{ij} \,\,=\,\, \rm det\,\, \it f_{ij} \,\,;$ and  $w^{ij}_i
\,\,=\,\,0 $, which is equivalent to Dirac's gauge. 
In order to write the matrices of the system in a simple way, the following
auxiliary quantities are defined: 
\bea
	q^{ij}&:=&\beta^i\beta^j-N^2\psi^{-4}\tilde{\gamma}^{ij}, \\
	Q^i&:=&\left(\begin{array}{lll} q^{1i} & q^{2i} & q^{3i} \end{array}\right), \\
	-\delta^i&:=&\left(\begin{tabular}{c}
             $-\delta^i_1$ \\
             $-\delta^i_2$ \\
             $-\delta^i_3$ \\
           \end{tabular}\right).
\eea
Then, the explicit form of the matrices $\mathrm{ {\bf A}}^l$ are:
\bea
\mathrm{ {\bf A}}^l =
\left(
\begin{tabular}{c|c}
$0_{6\times6}$   & $0_{6\times24}$ \\ \hline
$0_{24\times6}$  & \begin{tabular}{c|c}
                   $-2\beta^lI_6$ & \begin{tabular}{ccc}
                                    $Q^l$ &          & 0     \\
                                          & $\ddots$ &       \\
                                    0     &          & $Q^l$ \\
                                     \end{tabular}             \\ \hline
                    \begin{tabular}{c|c}
                    $-\delta^l$ & $0_{3\times5}$ \\ \hline
                    $0_{15}$    & \begin{tabular}{c|c}
                                  $-\delta^l$ & $0_{3\times4}$ \\ \hline
                                  $0_{12}$    & \begin{tabular}{c|c}
                                                $-\delta^l$ & $0_{3\times3}$ \\ \hline
                                                $0_9$       & \begin{tabular}{c|c}
                                                              $-\delta^l$ & $0_{3\times2}$ \\ \hline
                                                              $0_6$       & \begin{tabular}{c|c}
                                                                            $-\delta^l$ & $0_3$ \\ \hline
                                                                            $0_3$   & $-\delta^l$ \\
                                                                             \end{tabular} \\
                                                              \end{tabular} \\
                                                \end{tabular}  \\
                                  \end{tabular}  \\
                    \end{tabular}     & $0_{18\times18}$ \\
                   \end{tabular}
\end{tabular}
\right)
\eea
 \\
\\

\section{Characteristic structure of the reduced system}
\label{s:characteristics}
Let us present here a preliminary analysis of the mathematical structure of 
system $\left(\ref{sisth1}\right)$.

First, we give the explicit expressions of the characteristic speeds in terms of the 
functions $\psi$, $N$, $\beta^i$ and $\tilde{\gamma}^{ij}$.

\noindent {\em Lemma 1:\,\, Let us consider the evolution vector $\partial_t$, whose 
components are $\xi^{\alpha}=\left(1,0,0,0\right)$,
and a generic spacelike covector of components 
$\zeta_{\alpha}=\left(0,\zeta_i\right)$ 
orthogonal to the evolution vector.
The associated eigenvalue problem (see, e.g., ref. \cite{Anile}):
\be
\label{e:eigen}
	\left[\mathrm{{\bf A}}^l\zeta_l-\lambda \mathrm{{\bf I}}\right]\mathrm{{\bf X}}_{\lambda}=0,
\ee
\noindent
where $\lambda$ denotes the eigenvalue and $\mathrm{{\bf X}}_{\lambda}$ the corresponding 
eigenvector, has the following solution:
\bea
  \lambda_0 &=& 0 , \nonumber \\
	\lambda_{\pm}^{\left(\zeta\right)} &=& -\beta^{\mu}\zeta_{\mu} \pm
	\frac{N}{\psi^2}\left(\tilde{\gamma}^{\mu\nu}\zeta_{\mu}
	\zeta_{\nu}\right)^{1/2} \nonumber \\
	&=& - \beta^{\mu}\zeta_{\mu} \pm N\left(\zeta^{\mu}\zeta_{\mu}\right)^{1/2},
  \label{eigenvalues}
\eea
where $\lambda_0$ has multiplicity 18, and each 
$\lambda_{\pm}^{\left(\zeta\right)}$ has multiplicity 6.
} 

Imposing Dirac's gauge in (\ref{e:gauges}) indeed guarantees the real
character of the eigenvalues corresponding to matrices $\mathrm{{\bf A}}^i$, and therefore
the hyperbolicity of the evolution system.  Even though this is not a
prerogative of the Dirac gauge, other prescriptions for $H^i$ in condition
(\ref{e:gauges}) lead to a more complicated structure of the resulting
sources. As mentioned after Eq. (\ref{e:shift}), a more important point is the
fact that other choices of $H^i$ will generally introduce time derivatives of
$h^{ij}$ in the elliptic subsystem, complicating further the complete PDE
system.  Of course, if no gauge is imposed at all, one can check that the
$\mathrm{{\bf A}}^l$ matrices admit complex eigenvalues. This reflects the property that
Einstein equations by themselves do not have a definite type, without the
specification of a gauge.  We conclude that when imposing Dirac's gauge the
eigenvalues of the linear combination $\mathrm{{\bf A}}^l\zeta_l$ are real:

\noindent {\em Lemma 2:\,\, Dirac's gauge is a sufficient condition 
for the hyperbolicity of 
system $\left(\ref{sisth1}\right)$}.

In the above eigenvalue problem, 
the first 6 eigenvectors, with 0 eigenvalue and associated with the $h^{ij}$ components
of $\mathrm{{\bf\bar{v}}}$ in (\ref{bar_v}), completely decouple from the other eigenvectors.
Therefore, the rest of eigenvectors can be studied independently.  For the
sake of clarity in the notation, let us define some auxiliary
quantities before writing the matrix of (right-)eigenvectors:
\bea
\mathrm{{\bf C}_1} & := & \left(\begin{array}{cc}
 -\zeta_iq^{i2} & -\zeta_iq^{i3} \nonumber \\
 \zeta_iq^{i1}  & $0$ \\
 $0$            & \zeta_iq^{i1}
\end{array}\right) \\
\mathrm{{\bf C}_2} & := & \left(\begin{array}{c} \zeta_1 \\ \zeta_2 \\ \zeta_3 \end{array}\right) \ \ .
\eea

The matrix of (right) eigenvectors, $\mathrm{{\bf R}^{(\zeta)}}$, associated with the eigenvalue 
problem described in the above Lemma 1 is:
\bea
\mathrm{{\bf R}^{(\zeta)}}=
\left(
\begin{tabular}{c|c}
$I_6$            & $0_{6\times24}$ \\ \hline
$0_{24\times6}$  & \begin{tabular}{c|c|c}
                      $0_{6\times12}$  & $-\lambda_+^{\left(\zeta\right)}I_6$ & $-\lambda_-^{\left(\zeta\right)}I_6$ \\ \hline
                      \begin{tabular}{cccc}
                         $\mathrm{{\bf C}_1}$ &       &          &  $0$ \\
                               & $\mathrm{{\bf C}_1}$ &          &  \\
                               &       & $\ddots$ &  \\
                         $0$   &       &          &  $\mathrm{{\bf C}_1}$ \\
                      \end{tabular}                                           &
                      \begin{tabular}{cccc}
                         $\mathrm{{\bf C}_2}$ &       &          &  $0$ \\
                               & $\mathrm{{\bf C}_2}$ &          &  \\
                               &       & $\ddots$ &  \\
                         $0$   &       &          &  $\mathrm{{\bf C}_2}$ \\ 
                      \end{tabular}                                           &
                      \begin{tabular}{cccc}
                         $\mathrm{{\bf C}_2}$ &       &          &  $0$ \\
                               & $\mathrm{{\bf C}_2}$ &          &  \\
                               &       & $\ddots$ &  \\
                         $0$   &       &          &  $\mathrm{{\bf C}_2}$ \\ 
                      \end{tabular} \\
                   \end{tabular}
\end{tabular}
\right)
\eea
If the determinant of this matrix vanishes, the set of eigenvalues
is not complete. This happens in the following cases:
\begin{enumerate}
	\item[-] {\em Case 1}: $\lambda_+^{\left(\zeta\right)}=\lambda_-^{\left(\zeta\right)}$. Since
	  \bea
\lambda_+^{\left(\zeta\right)}=\lambda_-^{\left(\zeta\right)}
\,\,\Rightarrow \,\,
N^2\psi^{-4}\zeta_i\zeta_j\tilde{\gamma}^{ij}=N^2\zeta_i\zeta^i=0 \ \ ,
\eea
and $\zeta_i\zeta^i$ does not vanish ($\zeta^i$ is a spatial vector different 
from zero) non-completeness only occurs if the lapse $N$ vanishes.

	\item[-] {\em Case 2}: $\zeta_i\zeta_jq^{ij}=0$. From the definition of $q^{ij}$, it follows
\bea
\label{e:basis-2}
\zeta_i\zeta_j\left(\beta^i\beta^j-N^2\psi^{-4}\tilde{\gamma}^{ij}\right)=0 
\nonumber
\\
\Leftrightarrow 
\left(\zeta_i\beta^i\right)^2=N^2\left(\zeta_i\zeta^i\right).
\eea 
One can see that the previous equality depends only on the direction of the vector $\zeta^i$ (i.e. $\zeta^i \zeta_ i =1$). From now up to the end of the study of the different cases, the vector $\zeta^i$ will be considered to be unitary. So $(\ref{e:basis-2})$ leads to:
\be
\label{e:basis-2bis}
\zeta_i\zeta_j\left(\beta^i\beta^j-N^2\psi^{-4}\tilde{\gamma}^{ij}\right)=0 
\nonumber
\\
\Leftrightarrow \left(\zeta_i\beta^i\right)^2=N^2.
\ee
Decomposing $\beta^i$ into components parallel and normal
to $\zeta^i$, we write $\beta^i= \left(\beta^\parallel\right)\zeta^i+\left(\beta^\bot\right)^i$,
where $\left(\beta^\parallel\right)= \zeta_i\beta^i$ and 
$\zeta_i\left(\beta^\bot\right)^i=0$.
From (\ref{e:basis-2bis}), we conclude:
\be
\label{e:basis-2a}
\zeta_i\zeta_j q^{ij}=0 
\Leftrightarrow \\
\left(\beta^\parallel\right)^2=N^2 \ \ .
\ee
Note that this case is independent of the choice of
$\zeta^i$, since it corresponds to $\left(\beta^\parallel\right)^i\left(\beta^\parallel\right)_i$,
i.e. $|\zeta^i \beta_i|^2$.
Therefore, non-completeness occurs if $|\beta^\parallel|=N$.

\item[-] {\em Case 3}:	$\zeta_iq^{ij}=0$, $\forall j=1,2,3$. This is a stronger case
than the previous one. Again from the definition of $q^{ij}$, we have:
\be
\label{e:basis-3}
\zeta_i\left(\beta^i\beta^j-N^2\psi^{-4}\tilde{\gamma}^{ij}\right)=0
\Leftrightarrow
\left(\zeta_i\beta^i\right)\beta^j =N^2\zeta^j.
\ee
From this, and the decomposition $\beta^i=\left(\beta^\parallel\right)\zeta^i+\left(\beta^\bot\right)^i$, it follows:
\be
\zeta_iq^{ij}=0
\Leftrightarrow \left(\beta^\bot\right)^i = 0 \ \ \hbox{ and } \ \ 
\left(\beta^\parallel\right)^2=N^2 
\ee
This is just a stronger version of the second case above.

\end{enumerate}
As a consequence of the above analysis we can set up the following lemma.

\noindent {\em Lemma 3:\,\,  
The (right-)eigenvectors associated with the matrix 
$\mathrm{{\bf A}}^l\zeta_l$  define a complete system
iff i) the lapse $N$ does not vanish, and ii) the projection of the evolution vector onto the plane spanned by 
$n^\mu$ and $\zeta^\mu$, i.e. $\left(t^\parallel\right)^\mu = N n^\mu + 
b_\beta \zeta^\mu$, is non-null, i.e. $\left(\beta^\parallel\right)^2\neq N^2$.}

In the eigenvalue problem (\ref{e:eigen}), $\zeta^i$ stands for an arbitrary
spatial vector. In particular, we can always choose $\zeta^i = \beta^i$. In that case,
the degeneracy condition in cases 2 and 3 above reduces to $\beta^i \beta_i = N^2$. This happens
if the vector $t^\mu$ becomes null. Moreover, if the vector $t^\mu$ is spacelike then 
we are in case 2, since then there exists
a vector $\zeta^i$ (in fact, a cone obtained by the rotation of the non-vanishing $\beta^i$
by an appropriate angle) 
such that the projection of $\beta^i$ onto that $\zeta^i$, refered to as $\left(\beta^\parallel\right)^i$,
satisfies $\left(\beta^\parallel\right)^i\left(\beta^\parallel\right)_i = 
\left( \zeta_ i \beta^i \right)^2 = N^2$.
We conclude:

\noindent {\em Proposition 1 : \, \,
The system (\ref{sisth1}) is {\em strongly hyperbolic} if $t^\mu$ is timelike, i.e.
if $N\neq 0$ and $N^2 - \beta^i\beta_i > 0$.
}

\medskip 

In some particular cases, degeneracy in the eigenvalues can occur. In particular,
it could happen that one of the eigenvalues $\lambda_+$ or $\lambda_-$ coincides with
$\lambda_0$. These degeneracies can appear where: 
\be
\label{e:degeneracies}
\lambda_+^{\left(\zeta\right)}\lambda_-^{\left(\zeta\right)}=0 \Leftrightarrow \left(\beta^{\mu}\zeta_{\mu}\right)^2=N^2\left(\zeta^{\mu}\zeta_{\mu}\right).
\ee
Again, one can consider $\zeta^i$ to be unitary. Hence, either $\lambda_+$ or $\lambda_-$ vanishes when $\left(\beta^\parallel\right)^2 = N^2$. As seen in (\ref{e:basis-2bis}),
in this case the system of eigenvectors is incomplete.

Another relevant property is the following:

\noindent {\em Proposition 2:\,\, 
All the characteristic fields associated with the eigenvalue problem (\ref{e:eigen})
are {\em linearly degenerate}, i.e., they satisfy the following condition:
\be
\label{lindeg}
	{\bf D}\lambda_p\left(\bar{v}\right)\cdot r_p\left(\bar{v}\right) \,\,=\,\,0,
\ee
where $r_p$ is the eigenvector associated to the eigenvalue $\lambda_p$, 
and the operator ${\bf D}$ is defined in the space of the variables of the system.} 

\noindent This shows the good behaviour of the Dirac gauge since,
in the language from fluid dynamics, it means that no shocks can be 
propagated along these curves, in particular gauge shocks. Hence, if there 
were discontinuities, they have to be contact discontinuities. 

Regarding the characteristics speeds $\lambda^{(\zeta)}_{\pm}$ we have:

\noindent {\em Corollary 1:\,\,
The non-zero eigenvalues associated with $\zeta^i$ correspond
to the coordinate velocity of light.
}

\noindent This feature, which is an expected result, 
can be shown by considering a unitary $\zeta^i$
and a curve whose spatial part 
points in the $\zeta^i$ direction: $\displaystyle{\frac{dx^i}{dt}= \left|\frac{dx^i}{dt}\right| \zeta^i}$. 
 Using the 3+1 expression of the metric, the vanishing of the line element
of the curve, where the component
of $\beta^i$ in the $\zeta^i$ direction is considered, is imposed. It follows,
using the expression for $\lambda^{(\zeta)}$ in (\ref{eigenvalues}) that 
$\displaystyle{\lambda^{(\zeta)}=\left|\frac{dx}{dt}\right|}$.

\subsection{Application to inner boundary conditions}
\label{s:inner_BCs}
The explicit expressions (\ref{eigenvalues}) for the characteristic speeds
are specially useful in the assessment of the boundary conditions to be imposed
on a given border.
We illustrate this by considering inner boundaries in the context of 
excised black hole spacetimes.
Before doing so, let us underline that the FCF can be employed in combination 
with any of the standard techniques dealing with the black hole singularity
in numerical evolutions of black hole spacetimes,
namely {\em excision}, {\em punctures} or {\em stuffed black holes}.
However, the excision technique is favoured if (the elliptic subsystem of)
the FCF is implemented by means of spectral methods.
Focusing on the excision approach, let us denote by ${\cal S}_t$ the inner sphere employed as inner boundary
at a given spacelike slice $\Sigma_t$, and by
${\cal H}$ the worldtube hypersurface generated along the evolution
by {\em piling up} the different  ${\cal S}_t$. 
A natural expectation is that no 
inner boundary conditions should be prescribed for radiation fields on inner
superluminal (growing) inner boundaries.  This would avoid the need to
incorporate boundary conditions in the well-posedness analysis of the
associated initial boundary value problem.  From this reason, spacelike inner
hypersurfaces $\cal{H}$ are good candidates for inner boundary conditions.
However, this general idea must be assessed in the context of every specific
evolution scheme.  In our particular case, we must check that characteristic
speeds (\ref{eigenvalues}) are outgoing (with respect to the integration
domain).  The tangent vector $h^\mu$ to ${\cal H}$ which is normal to each
${\cal S}_t$, and transports ${\cal S}_t$ into ${\cal S}_{t+\delta t}$, can be
written as 
\be
\label{e:hmu}
h^\mu = Nn^\mu + h_s s^\mu ,
\ee
where $s^\mu$ is the normal vector to ${\cal S}_t$, lying on $\Sigma_t$
and pointing toward spatial infinity. 
Then, since the norm of $h^\mu$ is given by $h^\mu h_\mu = -N^2 + h_s^2$, it follows that
${\cal H}$ is spacelike as long as $b>N$. Choosing a coordinate system 
adapted to ${\cal H}$, i.e. where all the spheres ${\cal S}_t$ stay at the same
coordinate position ---say $r= \mathrm{const}=r_o$--- it follows that $h_s= \beta^i s_i \equiv \beta^\perp$.
In this case, ${\cal H}$ is spacelike as long as $\beta^\perp>N$.
Evaluating expression  (\ref{eigenvalues}) for $\zeta^i = s^i$, it follows
\be
\begin{array}{ll}
 \lambda^{(s)}_{\pm} = -\beta^\perp \pm N 
 \end{array}
\ee
From this it follows that:

\noindent 
{\em Corollary 2:\,\,
For a coordinate system adapted to a spacelike inner
worldtube ${\cal H}$, where $\beta^\perp>N$, no ingoing radiative modes 
flow into the integration 
domain $\Sigma_t$ at the excision surface. } 

Under these conditions no inner boundary conditions whatsoever must be prescribed 
for the hyperbolic part. Of course, it is not obvious how to choose dynamically
an inner boundary ${\cal H}$ that is guaranteed to be
spacelike during the evolution.
A proposal in this line has been presented in \cite{JECI}
in the context of the dynamical trapping horizon framework (see e.g. Ref.
\cite{GourgJ06}). Quasi-local approaches to black hole horizons aim
at modeling the boundary of a black hole region as world-tubes of apparent
horizons $(\Sp_t)$. Dynamical horizons provide a geometric prescription
for ${\cal H}$ that is guaranteed to be spacelike, as long as the black hole
is dynamical, and remain inside the event horion, if cosmic censorship holds. 
The corresponding geometric dynamical horizon characterization
is enforced as an inner boundary condition on the  
the elliptic part of the FCF, in particular on the
shift equation (\ref{e:shift}).
This shows the key interplay between elliptic and 
hyperbolic modes in the coupled fully-constrained PDE evolution system.
Note however that, according to Proposition 1, the
hyperbolic evolution system ceases to be strongly hyperbolic. In fact, the evolution 
vector $t^\mu$, tangent to ${\cal H}$ in the adapted coordinate system, becomes
spacelike in a finite region. This can be bypassed by adopting a coordinate
system in which the coordinate radii of the ${\cal S}_t$ slices grow  in time:
$r = r(t) \neq 0$, where $r(t)$ is appropriately {\em chosen}. 
In this case, $h_s= \beta^\perp$ holds no longer, and this
relation is rather substituted by $\beta^\perp = h_s - \left[r(t) -r_o\right]$. 
This condition is again under {\em  control} through the appropriate boundary condition 
on the elliptic equation for $\beta^i$. Note that in this case the characteristics are
still {\em outgoing} from the integration domain though, in this case with a coordinate
growing excision sphere, this feature is no longer
characterized by the negativity of the characteristics speeds $\lambda^{(s)}_{\pm}$.
The outgoing character is guaranteed by the characterization of $\lambda^{(s)}_{\pm}$
as the coordinate velocity of light in {\em Corollary 1}, together with the spacelike character of ${\cal H}$.

\section{Dirac gauge and system of conservation laws}
\label{s:conservation_laws}
A hyperbolic system of conservation laws, without sources, is:
\be
\partial_t \mathrm{{\bf u}} + \mathrm{D}_i \mathrm{{\bf f}}^i(\mathrm{{\bf u})} = 0.
\ee
In this system we can identify the set of unknowns, i.e., the vector of conserved quantities $\mathrm{{\bf u}}$, and their corresponding fluxes $\mathrm{{\bf f}({\bf u})}$.

The choice of Dirac's gauge allows us 
to find the following set of $l$ vector fluxes $\mathrm{{\bf f}}^l$ ($l=1,2,3$), of dimension 30:
\bea
\label{e:flux}
	\mathrm{{\bf f}}^l : = \left( \begin{array}{c}
 \left(0_6\right) \\
 \left(-2u^{ij}\beta^l+w^{ij}_k\left[\beta^k\beta^l-N^2\psi^{-4}\tilde{\gamma}^{kl}\right]\right) \\
 \left(-u^{ij}\delta^l_k\right)
\end{array}\right).
\eea
\noindent
in terms of which 
system $\left(\ref{sisth1}\right)$ can be rewritten as a 
{\it hyperbolic system of conservation laws (with sources)}.
The Jacobian matrices associated
to the fluxes $\mathrm{{\bf f}}^l$, $\left(\mathrm{{\bf A}}^*\right)^l$ are:\\ \\
\bea
\left(\mathrm{{\bf A}}^*\right)^l=
\left(
\begin{tabular}{c}
$0_{6\times30}$ \\ \hline
\begin{tabular}{c|c|c}
$-N^2\psi^{-4}E^l$ &  $-2\beta^lI_6$ & \begin{tabular}{ccc}
                                      $Q^l$ &          & 0     \\
                                            & $\ddots$ &       \\
                                      0     &          & $Q^l$ \\
                                     \end{tabular}             \\ \hline
$0_{18\times6}$  &  \begin{tabular}{c|c}
                     $-\delta$ & $0_{3\times5}$ \\ \hline
                     $0_{15}$  & \begin{tabular}{c|c}
                                 $-\delta^l$ & $0_{3\times4}$ \\ \hline
                                 $0_{12}$    & \begin{tabular}{c|c}
                                               $-\delta^l$ & $0_{3\times3}$ \\ \hline
                                               $0_9$       & \begin{tabular}{c|c}
                                                             $-\delta^l$ & $0_{3\times2}$ \\ \hline
                                                             $0_6$       & \begin{tabular}{c|c}
                                                                           $-\delta^l$ & $0_3$ \\ \hline
                                                                           $0_3$   & $-\delta^l$ \\
                                                                           \end{tabular} \\
                                                             \end{tabular} \\
                                               \end{tabular}  \\
                                 \end{tabular}  \\
                    \end{tabular}     & $0_{18\times18}$ \\
                   \end{tabular}
\end{tabular}
\right) \ \ ,
\eea 
where
\bea
&E^{ij,l}&:= \left(\begin{array}{cccccc}
	w^{ij}_1\delta^l_1 & w^{ij}_{(1}\delta^l_{2)} & w^{ij}_{(1}\delta^l_{3)} & w^{ij}_2\delta^l_2 & w^{ij}_{(2}\delta^l_{3)} & w^{ij}_3\delta^l_3 \end{array}\right), \nonumber \\
&E^l&:= \left(\begin{tabular}{c}
	              $E^{11,l}$ \\
	              $E^{12,l}$ \\
	              $E^{13,l}$ \\
	              $E^{22,l}$ \\
	              $E^{23,l}$ \\
	              $E^{33,l}$ \\
		          \end{tabular}\right),
\eea
and the {\em parentheses} in the subindices represent 
a symmetric sum, e.g., $w^{ij}_{(1}\delta^l_{2)} = w^{ij}_{1}\delta^l_{2} + w^{ij}_{2}\delta^l_{1}$.

These matrices have the same eigenvalues as
the matrices $\mathrm{{\bf A}}^l$. The corresponding eigenvectors are
different but they keep the same fundamental properties as the ones
associated to the matrices $\mathrm{{\bf A}}^l$, namely they define a complete system.
Hence, the following lemma is in order:

{\em Proposition 3:\,\, Taking advantage of Dirac's gauge, it is possible to
convert the hyperbolic part of the coupled elliptic-hyperbolic system of
the FCF formalism, into a (strongly) hyperbolic system of conservation 
laws (with sources)}.

\section{Preservation of the Dirac gauge in the evolution: the Dirac system}
\label{s:Dirac_system}

The importance of the enforcement of the Dirac gauge during the evolution in time has
already been stressed in the introduction. In this section we give a
brief description of some numerical algorithms that can be used to fulfill the Dirac
gauge, when solving the reduced system (\ref{e:hyperbolic_part}). 
In particular, we do not intend to provide a formal proof of the consistency of
the method.
Because of
the unimodularity of the conformal metric $\tilde{\gamma}_{ij}$, the symmetric
tensor $h^{ij}$ has only five degrees of freedom. For simplicity, here we shall
illustrate the scheme by considering the case where the trace $h = f_{ij}h^{ij} = 0$. The unimodular
condition would be satisfied by an iteration on the value of the trace, as
described in \cite{BonazGGN04}. We consider the particular case of spherical
polar coordinate system $(r,
\theta, \varphi)$, and note by $\Delta$ the flat Laplace operator, i.e.
\begin{equation}
  \label{e:def_lap}
  \Delta := \DSc_i\DSc^i = \dder{}{r} + \frac{2}{r} \der{}{r} + \frac{1}{r^2}\lapang  \ \ , 
\end{equation}
where $\lapang$ involves only angular derivatives. Thus,
the problem to be solved can be written as a wave equation with constraints 
\begin{eqnarray}
  \left(\dder{}{t} - \Delta \right) h^{ij} &=& \Sp^{ij} \ \ , \label{e:wave_tensor} \\
  \DSc_j h^{ij} &=& 0 \ \ , \label{e:Dirac_gauge} \\
  h &=& 0 \ \ ; \label{e:trace_zero}
\end{eqnarray}
where the source $\Sp^{ij}$ gathers all the other terms of
Eqs.~(\ref{e:hyperbolic_part_detailed}), including the shift terms
in the differential operator. The structure of the differential
operator in the left-hand side is here simplified with respect to the full
evolution one of Sec.~\ref{s:first_order_reduction}, in order to focus on the
propagation aspects, which are already contained in the simple wave operator.
The full evolution operator can also be handled with a similar technique, but
involving more technical justifications. The
system~(\ref{e:wave_tensor})-(\ref{e:trace_zero}) can be seen as the evolution
of two scalar fields, two dynamical degrees of freedom, from which one
recovers the full tensor $h^{ij}$ using the trace and divergence-free
conditions. To gain insight, it is helpful to decompose the tensor on a basis
of Mathews-Zerilli~\cite{Mathe62, Zeril70} tensorial spherical harmonics. We
use the basis of six families of {\em pure-spin tensor harmonics\/} as referred to
by Thorne~\cite{Thorn80}, with the same notations: $\w{T}^{L_0,\ell m},
\w{T}^{E_1,\ell m}, \w{T}^{B_1,\ell m}, \w{T}^{E_2,\ell m}, \w{T}^{B_2,\ell
  m}, \w{T}^{T_0,\ell m}$. If we note the coefficients of $h^{ij}$ in this
basis $\left( c^{L_0,\ell m}, c^{E_1,\ell m}, c^{B_1,\ell m}, c^{E_2,\ell m},
  c^{B_2,\ell m}, c^{T_0,\ell m} \right)$, we can define for any rank~2
symmetric tensor the following six scalar fields:
\begin{eqnarray}
  L_0 &:=& \sum_{\ell, m} c^{L_0, \ell m} Y_{\ell m} = h^{rr} \nonumber,\\
  \eta &:=& \sum_{\ell\geq 1, m} c^{E_1, \ell m} Y_{\ell m} \nonumber,\\
  \mu &:=& \sum_{\ell\geq 1, m} c^{B_1, \ell m} Y_{\ell m} \nonumber,\\
  {\cal W} &:=& \sum_{\ell\geq 2, m} c^{E_2, \ell m} Y_{\ell m} \nonumber,\\
  {\cal X} &:=& \sum_{\ell\geq 2, m} c^{B_2, \ell m} Y_{\ell m} \nonumber,\\
  T_0 &:=& \sum_{\ell, m} c^{T_0, \ell m} Y_{\ell m} \label{e:def_potentials} \ \ ,  
\end{eqnarray}
where $Y_{\ell m}(\theta, \varphi)$ are the scalar spherical harmonics, which
are eigenfunctions of the angular Laplace operator $\lapang Y_{\ell m} =
-\ell(\ell+1) Y_{\ell m}$. Note that there is a one-to-one relation between
the six components of $h^{ij}$ and these six scalar fields. The trace
condition~(\ref{e:trace_zero}) simply turns into $T_0 + h^{rr} = 0$, therefore
we shall replace $T_0$ with $-h^{rr}$ in all forthcoming expressions. The
divergence-free conditions~(\ref{e:Dirac_gauge}) turn into:
\begin{eqnarray}
  \der{h^{rr}}{r} + \frac{3h^{rr}}{r} + \frac{1}{r} \lapang \eta  = 0,\label{e:dir1}\\ 
  \der{\eta}{r} + \frac{3\eta}{r} + \left( \lapang + 2 \right) \frac{{\cal W}}{r} -
  \frac{h^{rr}}{2r} = 0 ,\label{e:dir2}\\ 
  \der{\mu}{r} + \frac{3\mu}{r}+ \left( \lapang + 2 \right) \frac{{\cal X}}{r} = 0
  ;\label{e:dir3} 
\end{eqnarray}
where all the angular derivatives are expressed in terms of $\lapang$,
introduced in Eq.~(\ref{e:def_lap}). 

A first way to solve the system~(\ref{e:wave_tensor})-(\ref{e:trace_zero}) has
been described in Ref.~\cite{BonazGGN04} and uses evolution equations for
$h^{rr}$ and $\mu$, from which other scalar fields are deduced through the
gauge equations~(\ref{e:dir1})-(\ref{e:dir3}) as solutions of the angular
Laplace operator, with radial derivatives as sources. However, this method has
the great disadvantage of requiring the computation of two radial derivatives
to get $h^{ij}$, when the source $\Sp^{ij}$ already contains second-order
radial derivatives of $h^{ij}$. This fourth-order derivation introduces a
great amount of numerical noise, which has been observed to rapidly spoil the numerical
integration. An alternative way is to evolve two other scalar fields and then
to integrate (or solve PDEs coming from) the Dirac gauge condition to obtain
the others.  Unfortunately, this is not possible using only the six scalar
fields~(\ref{e:def_potentials}), but one can devise the following procedure in
a similar spirit.

Any rank~2 symmetric tensor $T^{ij}$can be split into two pieces:
\begin{equation}
  \label{e:helmholtz_tensor}
  T^{ij} = \left( \hat{L} V \right)^{ij} + \tilde{T}^{ij} \equiv \DSc^iV^j + \DSc^jV^i +
  \tilde{T}^{ij}, 
\end{equation}
with $\DSc_j\tilde{T}^{ij} = 0$. For a given $T^{ij}$ the divergence of
Eq.~(\ref{e:helmholtz_tensor}) allows for the determination of the vector
$V^i$ through the elliptic PDE
\begin{equation}
  \label{e:vect_poisson}
  \DSc^k\DSc_k V^i + \DSc^i\DSc_j V^j = \DSc_jT^{ij},
\end{equation}
where $V^i$ is fixed up to isometries of $f_{ij}$, which are set by the choice of
boundary conditions. If we now
return to the case $T^{ij} = h^{ij}$ and consider only asymptotically
flat spatial metric defined on $\mathbb{R}^3$ ---no holes--- the Dirac gauge
condition~(\ref{e:Dirac_gauge}) is equivalent to having $V^i=0$, since there are no
Euclidean symmetries vanishing at infinity. If one
similarly seeks three scalar fields $(A,B, C)$ such that:
\begin{equation}
  \label{e:prop_potentials}
  A=B=C=0 \iff \tilde{T}^{ij} = 0,
\end{equation}
one can check that a solution is:
\begin{eqnarray}
  A &=& \der{{\cal X}}{r} - \frac{\mu}{r}, \label{e:def_A}\\
  B &=& \der{{\cal W}}{r} - \frac{\lapang {\cal W}}{2r} - \frac{\eta}{r} - \frac{h^{rr}}{4r},
  \label{e:def_B}\\
  C &=& \der{h^{rr}}{r} + \frac{3h^{rr}}{r} + 2\lapang \left( \der{{\cal W}}{r} +
    \frac{{\cal W}}{r} \right). \label{e:def_C}
\end{eqnarray}
In the present case where the trace (or the determinant) is given, $B$ and $C$
are actually coupled and it is sufficient to consider:
\begin{eqnarray}
  \tilde{B} &=& \sum_{\ell, m} \tilde{B}^{\ell m} Y_{\ell m}, \text{ with
  }\nonumber\\
 \tilde{B}^{\ell m} &=& (\ell +2) \left( \der{{\cal W}}{r} + \ell
 \frac{{\cal W}}{r} \right) - \frac{2\eta}{r} -\frac{1}{2(\ell+1)} \left( \der{h^{rr}}{r} +
   (\ell+4)\frac{h^{rr}}{r} \right)   \label{e:def_tildeB},
\end{eqnarray}
to recover $B$ and $C$ using the trace. A nice property of $A$ and $\tilde{B}$
is that, when expressed in terms of these potentials related to $h^{ij}$, the
tensor Poisson equation, with $F^{ij}$ being a symmetric-tensor representing a source:
\begin{equation}
  \label{e:tensor_Poisson}
  \Delta h^{ij} = F^{ij}
\end{equation}
has a rather simple form. Namely, if we define $F^A$ and $F^{\tilde{B}}$ as the
scalar potentials similar to $A$ and $\tilde{B}$, but deduced from $F^{ij}$,
a consequence of Eq.~(\ref{e:tensor_Poisson}) is:
\begin{eqnarray}
  \Delta A &=& F^A, \nonumber\\
  \tilde{\Delta} \tilde{B} &=& F^{\tilde{B}} , \label{e:Poisson_AB} 
\end{eqnarray}
with
\begin{equation}
  \label{e:def_tilde_Laplace}
  \tilde{\Delta} := \dder{}{r} + \frac{2}{r} \der{}{r} + \frac{1}{r^2}
  \tilde{\Delta}_{\theta\varphi} \qquad \text{and} \qquad
  \tilde{\Delta}_{\theta\varphi} Y_{\ell m} := -\ell(\ell-1) Y_{\ell m}.
\end{equation}
Obviously, a very similar property holds for the wave
equation~(\ref{e:wave_tensor}). Therefore, a way of solving
numerically the constrained system of
Eqs.~(\ref{e:wave_tensor})-(\ref{e:trace_zero}), by making use of the
potentials $A$ and $B$, is the following. With the
source $\Sp^{ij}$ and $h^{ij}$ known at the initial hypersurface, it is
possible to deduce the potentials $\Sp^A$ and $\Sp^{\tilde{B}}$ of the source
and thus to advance the potentials $A$ and $\tilde{B}$ of $h^{ij}$ to next
time-step through the evolution equations
\begin{eqnarray}
  \left(\dder{}{t} - \Delta \right) A &=& \Sp^A, \nonumber\\
  \left(\dder{}{t} - \tilde{\Delta} \right) \tilde{B} &=& \Sp^{\tilde{B}}.
  \label{wave_potentials} 
\end{eqnarray}
Then the six scalar fields~(\ref{e:def_potentials}) can be computed by solving
the PDE system formed by the following five elliptic equations: the definitions of
$A$ and $\tilde{B}$, i.e. Eqs.~(\ref{e:def_A}) and (\ref{e:def_tildeB}), together with the
Dirac gauge conditions (\ref{e:dir1})-(\ref{e:dir3}) plus the trace-free
condition~(\ref{e:trace_zero}) ---used to get $T_0$. All the components of $h^{ij}$
can be finally recovered by taking angular derivatives of the scalar fields defined in
Eqs.~(\ref{e:def_potentials}). With this algorithm, only two scalar potentials, $A$ and $\tilde{B}$, are evolved in time.  The whole tensor is
deduced from these potentials and the gauge and trace conditions. Note that, when decomposing all
the scalar fields onto a spherical harmonics function basis, the elliptic system of
five PDEs described above reduces to a system of coupled {\em ordinary\/}
differential equations in the radial coordinate $r$.

With either of these approaches (the one described here or that presented in Ref. 
\cite{BonazGGN04}) it is possible to evolve two scalar
potentials using hyperbolic wave-like operators and recover the symmetric
tensor $h^{ij}$ through an elliptic system of PDEs obtained from the gauge
conditions. A numerical implementation of these techniques being beyond the
scope of the present article, we have here only exhibited both algorithms in
order to show that it is, in principle, possible to build-up the whole
conformal metric from the gauge conditions, while being consistent with the
evolution equations. This might inversely be linked toward the property of the
Dirac gauge system being preserved by the 3+1 evolution system. Future
numerical developments in these directions shall certainly bring better
insight into the problem.

\section{Discussion.}
\label{s:discussion}
All evolution formalisms for the resolution of Einstein equations
as an initial value boundary problem exploit the intrinsic hyperbolicity
of Eqs. (\ref{e:EFE}), although the
associated evolution systems are not necessarily hyperbolic from
the PDE theory point of view \cite{Fri06}. 
In the present case of the FCF formalism \cite{BonazGGN04}, 
Einstein equations result in a coupled
elliptic-hyperbolic PDE system. 
The hyperbolic part PDE evolution system consists of the {\em reduced system}, governing the 
evolution of the gravitational degrees of freedom, whereas the elliptic part 
is formed by the {\em constrained system} and part of the {\em gauge system} 
(maximal slicing equation). In fact, 
in the context of the algorithms presented in section \ref{s:Dirac_system},
the elliptic Dirac system, Eqs. (\ref{e:dir1})-(\ref{e:dir3}), can be actually seen as a 
part of the PDE evolution system. 
In summary, the evolution PDE system is formed by the reduced, constraint, and gauge systems, whereas
the the fulfillment of the subsidiary system, represented by 
 Eq.~(91) in Ref. \cite{BonazGGN04} for $\dot{\beta^i}$, can be used as a control test
of the scheme along the evolution.
We have carried out a first analysis of the
mathematical structure of the PDE evolution system paying particular attention to the
equations (\ref{e:hyperbolic_part}) governing the evolution for the deviation
$h^{ij}$ of the conformal metric from the flat fiducial one $f_{ij}$, i.e.
$h^{ij} = \tilde{\gamma}^{ij} - f^{ij}$. 
Dirac's gauge plays an important role in getting a well defined hyperbolic structure.
This elliptic gauge is close in spirit and properties to other gauges employed in the
literature, like the {\em spatial harmonic} gauge in \cite{AndMon03},
the {\em minimal distortion} introduced by York \& Smarr, the {\em new minimal distortion}
gauge introduced by Jantzen \& York, or the 
numerically motivated {\em pseudo-minimal distortion} gauge by Nakamura, 
{\em approximate minimal distortion} by Shibata or the {\em Gamma freezing}
(cf. Secs. 9.3. and 9.4 in  Ref. \cite{Gou07a} for
a review of them). In particular, all of them can be written as
elliptic equations on the shift vector $\beta^i$. The Dirac gauge fixes spatial coordinates in the
evolution (including on the initial data, as the {\em spatial harmonic gauge} does) up to 
boundary conditions.
For boundary conditions (enforced when solving the elliptic PDE for $\beta^i$)
such that the evolution vector is timelike,
the Dirac gauge provides a sufficient condition for the 
strong hyperbolicity of Eq. (\ref{e:hyperbolic_part}). Moreover,
using this gauge it is possible to derive a {\it flux vector}
in terms of which the first-order
system of equations, equivalent to (\ref{e:hyperbolic_part}), has the
structure of a hyperbolic system of conservation laws (with sources).
Likewise, the analysis of the characteristics sheds light on the 
prescription of inner boundary conditions on a spacelike   
inner cylinder, when employing an excision approach to black hole evolutions.
More generally, maximal and Dirac gauges can be relaxed to admit more
general gauges, while preserving the hyperbolic properties of the system 
but possibly complicating the structure of the sources.

Having said this, it is clear that further analysis is necessary. 
First, particular attention should be payed to the source terms in equation (\ref{e:hyperbolic_part_detailed}). They can introduce, in the so-called stiff case, new characteristic time scales (relaxation times in the language of fluid dynamics) which may be much smaller than the CFL (Courant-Friedrichs-Lewy) numerical time step (see, e.g., \cite{Jin00, Ross04, Mini07}). In particular, authors in reference \cite{Jin00} have studied general hyperbolic systems with supercharacteristic relaxations, and they shown in which conditions a source term can be damping or, on the contrary, enforces growth of instabilities. Looking, in our case, at the quantity $R_*^{ij}$ (Eq. (\ref{e:hyperbolic_part_detailed2})), one can notice the presence of quadratic terms in the $w^{ij}_k$; it suggests that huge spatial gradients of $h^{jk}$ can introduce some degree of stiffness in the source terms.
Second, nothing has been said about the possible outer boundary
conditions to be prescribed  when studying
the initial boundary value problem with an outer timelike cylinder.
Certainly, in this case the well-posedness analysis is more complicate.
However, thanks the 
enforcement of the constraint along the evolution,
there is no need of devising specific constraint preserving 
boundary conditions, and Sommerfeld-like conditions as in \cite{NovBon04, BucSar06} can
be straightforwardly employed. Third, nothing has been said
about the elliptic part and its coupling with the hyperbolic subsystem.
On the one hand, this coupling is crucial in the overall  well-posedness 
of the problem,
as clearly illustrated in the inner boundary conditions issue, where
inner boundary conditions on the elliptic part determine the ingoing
or outgoing nature of the characteristics in the hyperbolic part.
On the other hand, the analysis of the elliptic system by itself represents 
an outstanding challenge. This is illustrated by the  
XCTS elliptic system \cite{PfeYor03,Pfe05} referred to in Section \ref{ss:brief_review}, 
very closely related to the FCF
elliptic subsystem. We note that, in this case, no results on existence are available
and very little is known on uniqueness, where
recent numerical \cite{PfeYor05,JarAnsLim07} and analytical works \cite{BauOMuPfe07,Wal07} works
point toward the essential non-uniqueness of the system (related to a
wrong sign in the differential operator of the maximal slicing equation).
Fourth, nothing has been said about consequences on well-posedness of 
coupling matter equations to the gravitational degrees of freedom.

Although our analysis is far from being exhaustive, it has the advantage of
giving some clues about which numerical strategies are the most convenient in
order to solve Einstein equations in the FCF formalism.
In this sense, we have attempted to obtain some limited but concrete results, rather
than remained frozen by the  ``non-attainability'' of
complete and fully rigorous results.

\bigskip

\noindent{\em Acknowledgments}.

I.C-C. acknowledges a doctoral fellowship from the Spanish Ministerio de
Educaci\'on y Ciencia (MEC) (ref. AP2005-2857).  JLJ acknowledges the support
of the Marie Curie contract MERG-CT-2006-043501 in the 6th European Community
Framework Program. EG and JN acknowledge support from the ANR grant 06-2-134423 MATH-GR.
Work supported by the grant AYA2004-08067-C03-01 from the
MEC, and a France-Spain bilateral research grant (ref. HF2005-0115), as well
as the Hubert Curien exchange grant, from the French ministry for foreign
affairs.

\end{document}